# Role of carbon nanotube diameter on thermal interfacial resistance through the analysis of vibrational mismatch: A Molecular Dynamics approach

Ajinkya Sarode[a], Zeeshan Ahmed[a], Pratik Basarkar[a], Atul Bhargav[a], and Debjyoti Banerjee[b]

[a]Mechanical Engineering, Indian Institute of Technology Gandhinagar, India
[b]Department of Mechanical Engineering, Texas A&M University, US

**Abstract:**

Carbon nanotube (CNT) have been known to increase the heat transfer at the solid-liquid interfaces, but have a limitation due to the interfacial thermal resistance. Vibrational mismatch at the interface leads to this interfacial thermal resistance, which plays an important role in energy transfer at the boundary. Negligible work has been reported on the influence of CNT diameter on the resistance through the vibrational mismatch study. Molecular dynamics simulations have been performed to investigate the effect of CNT diameter on interfacial resistance between carbon nanotube (CNT) and water molecules. This work is an effort to understand the heat transfer phenomenon at the interface by quantifying the vibrational mismatch. Analysis of the vibrational spectra of CNT and water molecules is done to study the effect of CNT diameter on interfacial resistance. Starting with the initial configuration, and equilibrating the system of CNT and water molecules at 300 K and 1 atm, the CNT temperature is raised to 700 K by velocity rescaling. This system is now allowed to relax as a micro-canonical ensemble. Based on the lumped capacitance analysis, the time constant of the CNT temperature response is determined, which is then used to compute the interfacial thermal resistance.

The interfacial thermal resistance is observed to be relatively higher for the larger diameter nanotube. This is attributed to the higher vibrational mismatch existing for larger diameter CNT as a result of low overlapping region between vibrational density states of CNT and water molecules. For smaller diameter CNT, the interfacial thermal resistance is low which results in the efficient heat transfer at the interface thus, emphasizing the indispensable role of larger diameter CNTs in the cooling applications.

**Keywords:** Carbon Nanotube, Thermal interfacial resistance, Vibrational mismatch, Overlapping ratio.

## 1. Introduction:

In 1991, the discovery of the carbon nanotube (CNT) opened up fascinating avenues in scientific and technological research fields [1]. Because CNTs have exceptionally high thermal conductivities (estimated to be as high as 6600 W/m-K [2], 17 times that of bulk metallic copper), heat transfer applications are naturally among the most important applications for these materials. CNTs are also attractive in boiling applications, especially those in electronic cooling [3]. CNTs act as nanofins on the solid-fluid interface, thereby increasing heat transfer surface area. However, despite high thermal conductivity, the actual enhancement in the heat transfer phenomenon is not very significant. The reason behind this limitation is attributed to the thermal interfacial resistance between CNT and fluid molecules surrounding it. The existence of this resistance is because of the non-bonded interactions between the molecules viz. Vander Waals forces [4]. Similar effects have been reported in the literature. The contact resistance between CNT and silicon substrate is few orders smaller than the thermal interfacial resistance. Hence, this interfacial resistance has a significant effect on the performance of nanofins [5].

Extensive study has been going on the interfacial resistance between CNT and surrounding molecules in the past few years. It has been proven experimentally that thermal interfacial resistance exists in carbon nanotube suspensions [6]. As per the literature available, the values of the thermal interfacial resistance vary from $0.76 \times 10^{-8}$ m$^2$K/W to $20 \times 10^{-8}$ m$^2$K/W. The mismatch in the vibrational frequency between CNT and surrounding molecules leads to the interfacial resistance between them ultimately reducing the heat transfer at the interface. Therefore, it becomes necessary to study the vibrational mismatch between CNT and surrounding molecules to understand the existence of thermal interfacial resistance.

In this work, molecular dynamics (MD) simulation has been performed to study the effect of diameter of CNT on the thermal interfacial resistance between CNT and water molecules. The objective of this study is to understand the heat transfer phenomenon at the interface more precisely, through the analysis of the overlapping vibrational spectrum between CNT and water molecules. For the study, the independent MD simulations are performed on single walled carbon nanotube (SWCNT) having chirality as (5,5), (10,10), (15,15), (20,20) and (25,25) surrounded by water molecules using LAMMPS software [7]. The diameter of CNT depends on its chirality and is calculated as per the equation given in literature [8]. This paper is divided into parts consisting of simulation setup and procedure, followed by illustration of the obtained results with its discussion and conclusion at the end.

## 2. Simulation setup and procedure:

Molecular dynamics is a widely used tool for performing simulations at the molecular level and determining thermal, mechanical and other properties of interest. Verlet algorithm [9] is used to update the position and velocity of the atoms at each time step by integrating Newton's equation of motion. The potential interaction between the atoms is estimated through potential energy function (*V*) which is further used to calculate the force acting on them. This potential energy function depends on the position of individual atoms present in the simulation domain and is composed of bonded and non-bonded energy interactions given as follows,

$$V(r) = V_B(r) + V_{N-B}(r) \qquad (1)$$

The bonded interaction includes energy stored due to the bond-stretching, angle of bending and torsional contributions. CNT is modeled using AIREBO [10] potential because in addition to the above mentioned interactions, it also includes long range Van der Waals interactions. SPC/Fw model is used for water molecules since it has better structural and thermodynamic properties [11]. The non-bonded interactions are evaluated from the Van der Waals and the electrostatic (eg. Coulombic) interactions present within water molecules and between CNT and water molecules surrounding it.

$$V_{N-B} = V_{vdw} + V_{coul} \tag{2}$$

The Van der Waals potential is modeled by Lennard Jones (LJ) 12:6 potential and is defined as,

$$V_{vdw} = \sum_{i \neq j} 4\varepsilon_{ij} \left[ \left( \frac{\sigma}{r_{ij}} \right)^{12} - \left( \frac{\sigma}{r_{ij}} \right)^{6} \right] \tag{3}$$

where $\sigma$ and $\varepsilon$ are the LJ parameters. The electrostatic energy interaction is given by Coulombic potential as follows,

$$V_{coul} = \sum_{i \neq j} \frac{q_i q_j}{r_{ij}} \tag{4}$$

where $q_i$ and $q_j$ are the charge on atom $i$ and $j$ of atom pair $i, j$. The values of $\varepsilon_{ij}$ and $\sigma_{ij}$ for interaction between two different atom types are given by the Lorentz-Berthelot [12, 13] mixing rules as follows,

$$\varepsilon_{ij} = \sqrt{\varepsilon_i \varepsilon_j} \tag{5}$$

$$\sigma_{ij} = \frac{1}{2}(\sigma_i + \sigma_j) \tag{6}$$

The simulation domain consists of pre-calculated number of water molecules in a cube of side approximately 6 nm having periodic boundary conditions depending on the diameter of the CNT in order to maintain a bulk density of water as 996 kg/m³ at 1 atm pressure and 300 K temperature. The ratio of the number of water molecules ($N_w$) to the number of carbon atoms ($N_c$) in CNT has been kept constant for CNTs of different diameter in order to maintain the temperature rise of the surrounding water molecules approximately the same. Table 1 shows the various simulation parameters considered during the simulations.

**Table 1:** Parameters incorporated in simulations.

| CNT chirality | Nanotube diameter (nm) | Carbon atoms ($N_c$) | Water molecules ($N_w$) | Ratio ($N_w/N_c$) |
|---|---|---|---|---|
| (5,5) | 0.679 | 500 | 1800 | 3.60 |
| (10,10) | 1.356 | 1000 | 3600 | 3.60 |
| (15,15) | 2.034 | 1500 | 5400 | 3.60 |
| (20,20) | 2.713 | 2000 | 7200 | 3.60 |
| (25,25) | 3.391 | 2500 | 9000 | 3.60 |

To begin with, the initial configuration is obtained using Packmol [14] which randomly places the water molecules around the centrally located nanotube in the domain. Once the initial position is set, the system is allowed to minimize its potential energy. After minimization, the system is equilibrated as a micro canonical (NVE) ensemble. The system is then equilibrated to 300 K and 1 atm as an isothermal-isobaric (NPT) ensemble. The position and velocities of the atoms present in the domain are updated at each time step using Nose/Hover thermostat [15] and barostat [16]. Equilibration is followed by raising the temperature of the CNT to 700 K by directly rescaling the velocities of the carbon atoms. The velocity scaling is carried out as follows,

$$\frac{T_{req}}{T_{base}} = \left(\frac{v_{req}}{v_{base}}\right)^2 \tag{7}$$

where $T_{req}$ and $T_{base}$ is the required temperature of the CNT, while $v_{req}$ and $v_{base}$ is the velocity of the carbon atoms present in CNT [4]. The last step in the simulation is to allow the system to relax under constant energy during which the energy from the nanotube is transferred to the surrounding molecules. The temperature of the nanotube and water molecules is recorded at regular time intervals during the relaxation which is further used to calculate the thermal interfacial resistance. In addition to this, velocity auto-correlation function for the CNT and water molecules just surrounding the CNT is calculated on which Fast Fourier Transform (FFT) is performed to give the vibration spectrum. The velocity auto-correlation function is calculated as follows,

$$C(t) = <v(0).v(t)> \tag{8}$$

where *v(0)* and *v(t)* are the velocity of carbon atoms at *t*=0 and *t*=t.

## 3. Results and discussions:

Figure 1 (generated using Visual Molecular Dynamics [17]) shows the simulation box consisting of SWCNT placed at the center of the box surrounded by water molecules. The length of CNT is taken as 6 nm to neglect the effect of length on the thermal interfacial resistance [18] since, the value tends to saturate as the length of nanotube changes (increases).

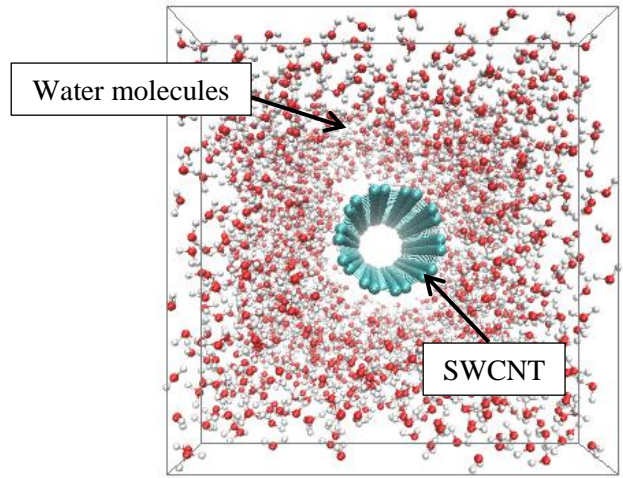

**Figure 1:** Simulation domain consisting of CNT (5,5) and water molecules

The temporal decay of nanotube temperature (and the increase in water temperature) during the relaxation period are plotted in Figure 2 for different diameters (chirality) of nanotube.

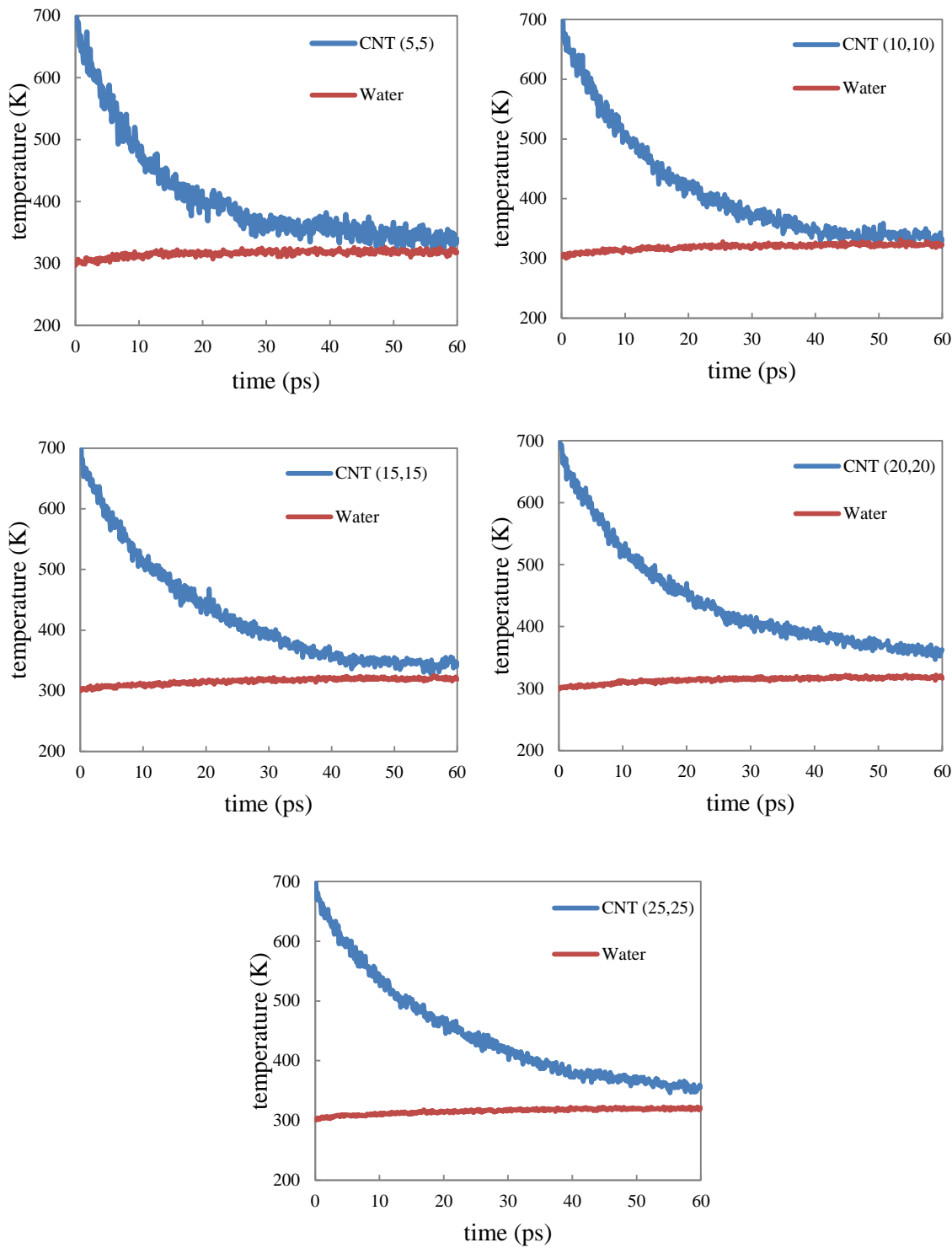

**Figure 2:** Temporal variation of CNT temperature and water molecules in the production run (NVE ensemble)

It can be seen that the temperature of the nanotube falls exponentially as it gives away the heat to the water molecules. Also, the temperature rise of water molecules at the end of the simulation is approximately the same in all the cases as expected. This is to ensure that the impact of surrounding temperature on the thermal interfacial resistance is of approximately same in all the simulations. Figure 3 shows the logarithmic decay (best fit) of the temperature difference between the nanotube and water molecules for all the cases.

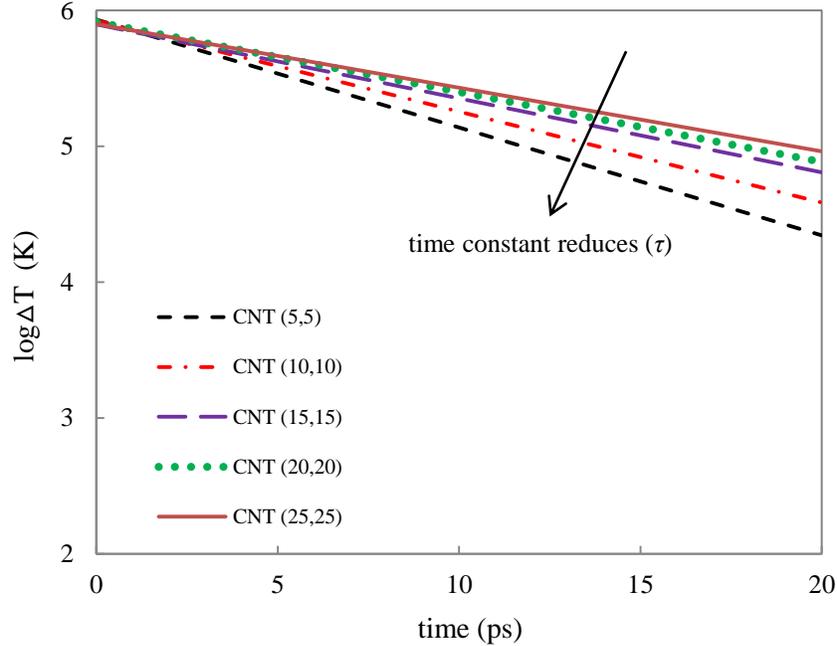

**Figure 3:** Logarithmic decay (best fit) of the temperature difference between CNT and water molecules in the production run (NVE ensemble)

As shown in the figure 3, the temperature decay of smaller diameter CNTs is faster than the larger diameter CNTs. The inverse of the slope of this logarithmic decay with respect to time gives the time constant which is later used in calculating the thermal interfacial resistance. Hence, the time constant for CNT (5,5) is the least while CNT (25,25) has the largest time constant. According to lumped capacitance analysis, the interfacial resistance at the nanotube-water interface can be calculated as [4],

$$\tau = \frac{mc_T R_k}{A_{CNT}} \quad (9)$$

where $\tau$ is the time constant, $m$ is the mass of the nanotube, $c_T$ and $A_{CNT}$ are the specific heat capacity and surface area of the nanotube, respectively while $R_k$ is the thermal interfacial resistance. The value of the specific heat capacity depends on the diameter of the nanotube [19], while the ratio of mass to the surface area is constant for a fixed length of the nanotube. Table 2 gives the time constant for different diameters of carbon nanotube. These values of the time constant are used to determine the thermal interfacial resistance between nanotube and water molecules at the interface and investigate the effect of diameter of CNT on it.

**Table 2:** Effect of nanotube diameter on the time constant.

| CNT chirality | nanotube diameter (nm) | time constant (ps) |
|---|---|---|
| (5,5) | 0.679 | 13.10 |
| (10,10) | 1.356 | 14.94 |
| (15,15) | 2.034 | 18.36 |
| (20,20) | 2.713 | 19.40 |
| (25,25) | 3.391 | 21.38 |

Figure 4 depicts the values of thermal interfacial resistance for different diameters of carbon nanotube. The resistance of $2.58\times10^{-8}$ ($m^2K/W$) for CNT of 5,5 chirality agrees well with the value obtained by Navdeep et al. [4]. Also, these values of resistance lie in the range of the interfacial resistance reported in the literature. It can be seen that there is a monotonous increase in the thermal interfacial resistance with the diameter of nanotube which can be attributed to the vibrational mismatch present at the interface.

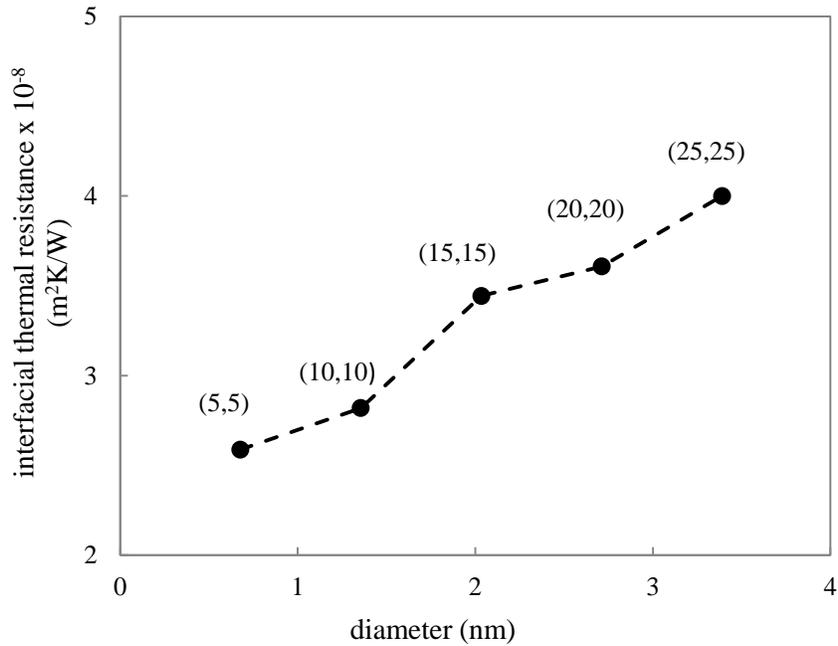

**Figure 4:** Effect of nanotube diameter on the thermal interfacial resistance

Therefore, to study the overlapping of the vibration density of states between CNT and water molecules, their vibration spectrums are analyzed. The low frequency vibration modes of CNT (radial breathing mode-RBM) play an important role in the heat transfer phenomenon at the interface [4] since, these vibrations nicely overlap with the librational modes of water molecules having frequency less than 30-40 THz [20]. Figure 5 shows the vibrational spectrum for carbon nanotube and water molecules just surrounding them.

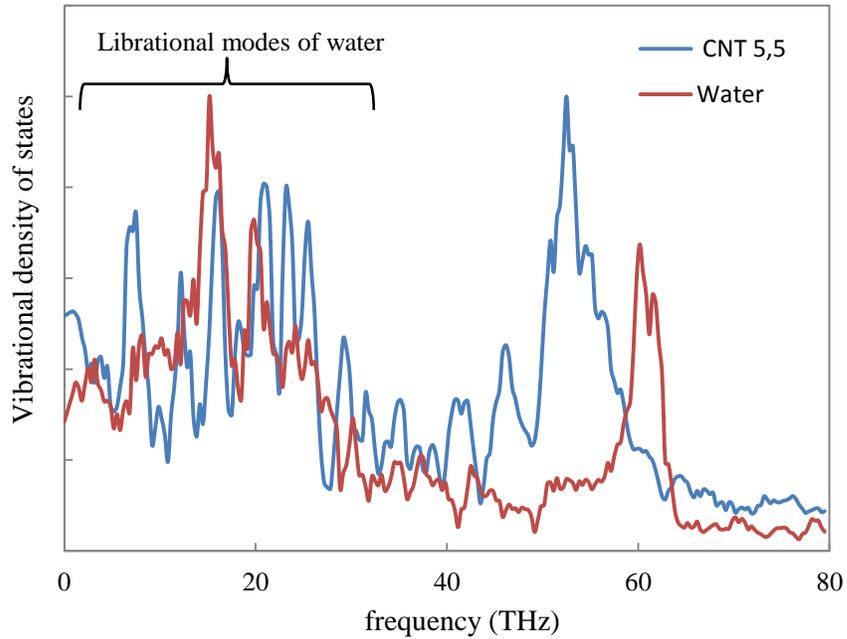

**Figure 5:** Vibrational density of states of carbon nanotube and water molecules just surrounding it

The peaks in the vibration spectrum are in good agreements with the fundamental vibrational frequencies of nanotube and water molecules [20,21]. As shown in figure 5, energy transfer between CNT and water molecules at the interface takes place mainly because of the overlapping of the vibration spectrum at the low frequency (libration) hence, a vibration range of 0-40THz is considered to study the overlapping ratio. Figure 6 shows the overlapping region between carbon nanotube and water molecules for different diameters of nanotube.

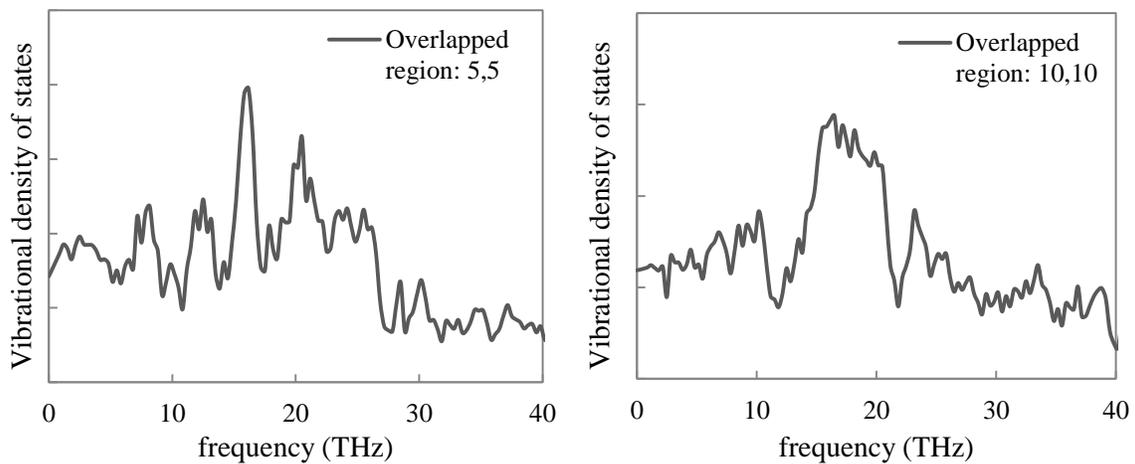

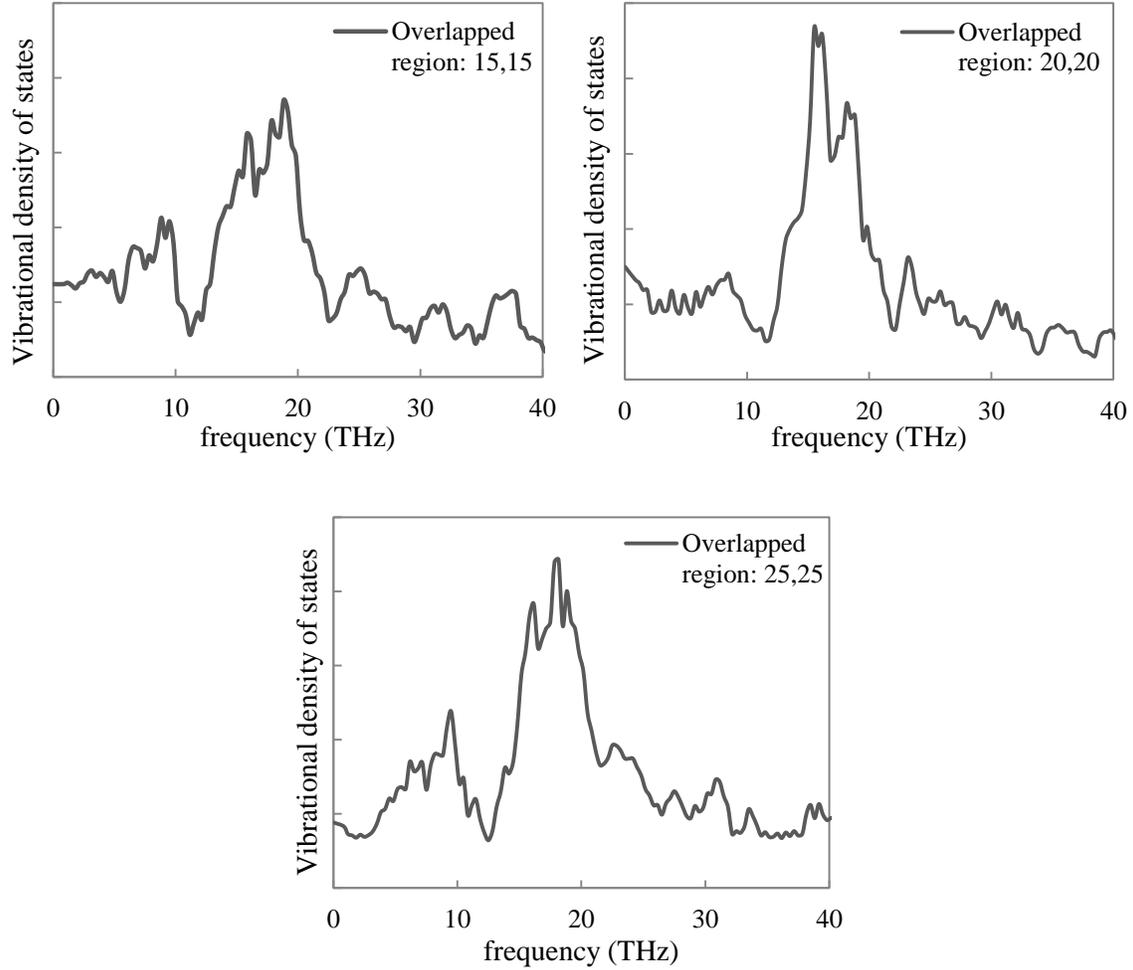

**Figure 6:** Vibrational density of states of overlapping region between carbon nanotube and water molecules just surrounding it

In order to quantify the overlapping regions for CNTs of different diameters, ratio of the area of overlapping region with respect to the total area of vibration spectrum of nanotube and water molecules is evaluated and termed as overlapping ratio. It is seen that as the overlapping ratio decreases due to rise in vibrational mismatch, the thermal interfacial resistance increases. Table 3 presents the complete results of the current work.

**Table 2:** Summarized simulation results showing the effect of diameter of nanotube

| CNT chirality | nanotube diameter $d$ (nm) | time constant $\tau$ (ps) | thermal interfacial resistance $R_k \times 10^{-8}$ ($m^2K/W$) | overlapping ratio |
|---|---|---|---|---|
| (5,5)   | 0.679 | 13.10 | 2.59 | 0.81 |
| (10,10) | 1.356 | 15.00 | 2.81 | 0.76 |
| (15,15) | 2.034 | 18.36 | 3.45 | 0.74 |
| (20,20) | 2.713 | 19.40 | 3.60 | 0.72 |
| (25,25) | 3.391 | 21.38 | 4.00 | 0.69 |

## 5. Conclusions:

Molecular dynamics simulations are performed to investigate the effect of diameter of SWCNT on the thermal interfacial resistance between nanotube and water molecules and study the effect with the help of an overlapping ratio. The resistance value obtained during the simulation of CNT (5,5) is in good agreement with the literature. Hence, the simulation is extended for nanotube of larger diameters. It is found that the value of thermal time constant of temperature decay increases with the increase in nanotube diameter. This leads to the rise in thermal interfacial resistance as the diameter of the nanotube increases. Analysis of vibrational spectrum of nanotube and surrounding water to quantify the vibrational mismatch explains the dependence of thermal interfacial resistance on the diameter of nanotube. Results show that larger diameter CNTs have more vibrational mismatch as compared to smaller diameter CNT. Thus, more vibrational mismatch makes the resonance between atoms/molecules (carbon and water molecules in this case) difficult at the interface which leads to poor coupling. This ultimately hinders the heat transfer at the interface resulting in higher value of thermal interfacial resistance. But, the rate at which resistance is varying with respect to diameter is not the same as the rate at which the overlapping ratio is changing. This further demands for a work to postulate a more precise relationship between thermal interfacial resistance and vibrational mismatch. To obtain a significantly high enhancement in the heat transfer, CNTs having smaller diameter should be preferred owing to its higher overlapping ratio and lower thermal interfacial resistance.